\documentclass[osajnl,twocolumn,showpacs,floatfix,bibnotes]{revtex4}

\usepackage{graphicx}

\newcommand\pictc[5]{\begin{figure}
                       \centerline{\vspace{0mm}
 \includegraphics[width=#1\columnwidth,height=0.7\textheight,keepaspectratio]{pict/#3}}
                       \protect\caption{\protect\label{fig:#4} #5}\vspace{0mm}
                    \end{figure}            }
\newcommand\pict[4][1]{\pictc{#1}{!tb}{#2}{#3}{#4}}
\newcommand\rpict[1]{\ref{fig:#1}}

\begin{document}
\begin{sloppy}

\title{Demonstration of all-optical beam steering in modulated photonic lattices}

\author{Christian R. Rosberg}
\author{Ivan L. Garanovich}
\author{Andrey A. Sukhorukov}
\author{Dragomir N. Neshev}
\author{Wieslaw Krolikowski}
\author{Yuri S. Kivshar}

\affiliation{Nonlinear Physics Centre and Laser Physics Centre,
Centre for Ultra-high bandwidth Devices for Optical Systems (CUDOS),
Research School of Physical Sciences and Engineering,
Australian National University,
Canberra ACT 0200, Australia}

\begin{abstract}
We demonstrate experimentally all-optical beam steering in modulated photonic lattices induced optically by three beam interference in a biased photorefractive crystal. We identify and characterize the key physical parameters governing the beam steering, and show that the spatial resolution can be enhanced by the additional effect of nonlinear beam self-localization.
\end{abstract}

\ocis{190.4420,190.5940,060.1810}

\maketitle

Periodic dielectric structures such as photonic crystals have attracted a lot of interest in recent years because they offer exciting possibilities to control and manipulate the propagation of light. In fact, a periodic modulation of the refractive index drastically modifies both the linear transmission spectrum and wave diffraction~\cite{Russell:1995-585:ConfinedElectrons}, and this strongly affects the linear as well as nonlinear propagation of optical beams. Furthermore, nonlinear self-phase modulation and localization of light in the form of spatial solitons~\cite{Kivshar:2003:OpticalSolitons} provide additional mechanisms for optical control of wave propagation~\cite{Morandotti:1999-2726:PRL, Meier:2005-1027:OL}
which is essential for applications such as all-optical signal routing and switching.

Dynamically reconfigurable and tunable photonic structures can be optically-induced in biased photorefractive crystals~\cite{Efremidis:2002-46602:PRE}, where nonlinear beam self-action and interaction can be observed at moderate laser powers. Such lattices were recently used to demonstrate many fundamental concepts of light propagation in one- and two-dimensional periodic structures, including generation of lattice and gap spatial solitons~\cite{Fleischer:2003-23902:PRL, Fleischer:2003-147:NAT, Neshev:2003-710:OL, Neshev:2004-83905:PRL}, nonlinear Bloch-wave interaction~\cite{Sukhorukov:2004-93901:PRL}, and tunable negative refraction~\cite{Rosberg:2005-2293:OL}.

One-dimensional optical lattices created by two interfering 
plane waves~\cite{Efremidis:2002-46602:PRE, Fleischer:2003-23902:PRL, Neshev:2003-710:OL} are inherently symmetric in the transverse direction ($x$) and invariant in the propagation direction ($z$). Recent theoretical studies~\cite{Kartashov:2005-1378:OL, Garanovich:2005-5704:OE} suggested the possibility of 
achieving 
controlled 
beam steering 
in such systems
by introducing a third interfering wave which breaks the lattice symmetry and induces modulation along both $x$ and $z$ directions. It was predicted that weak lattice modulation can induce a drift of broad solitons~\cite{Kartashov:2005-1378:OL}, and that binary switching can be obtained for strongly localized solitons~\cite{Garanovich:2005-5704:OE}. In this Letter, we study an experimentally accessible case of moderate lattice strength and modulation, where a probe beam extends over a few lattice sites. We present the first experimental demonstration of optically-controlled beam steering in modulated lattices, and show that simultaneous beam self-focusing allows to compensate for diffraction and obtain increased spatial resolution for practical applications.

\pict{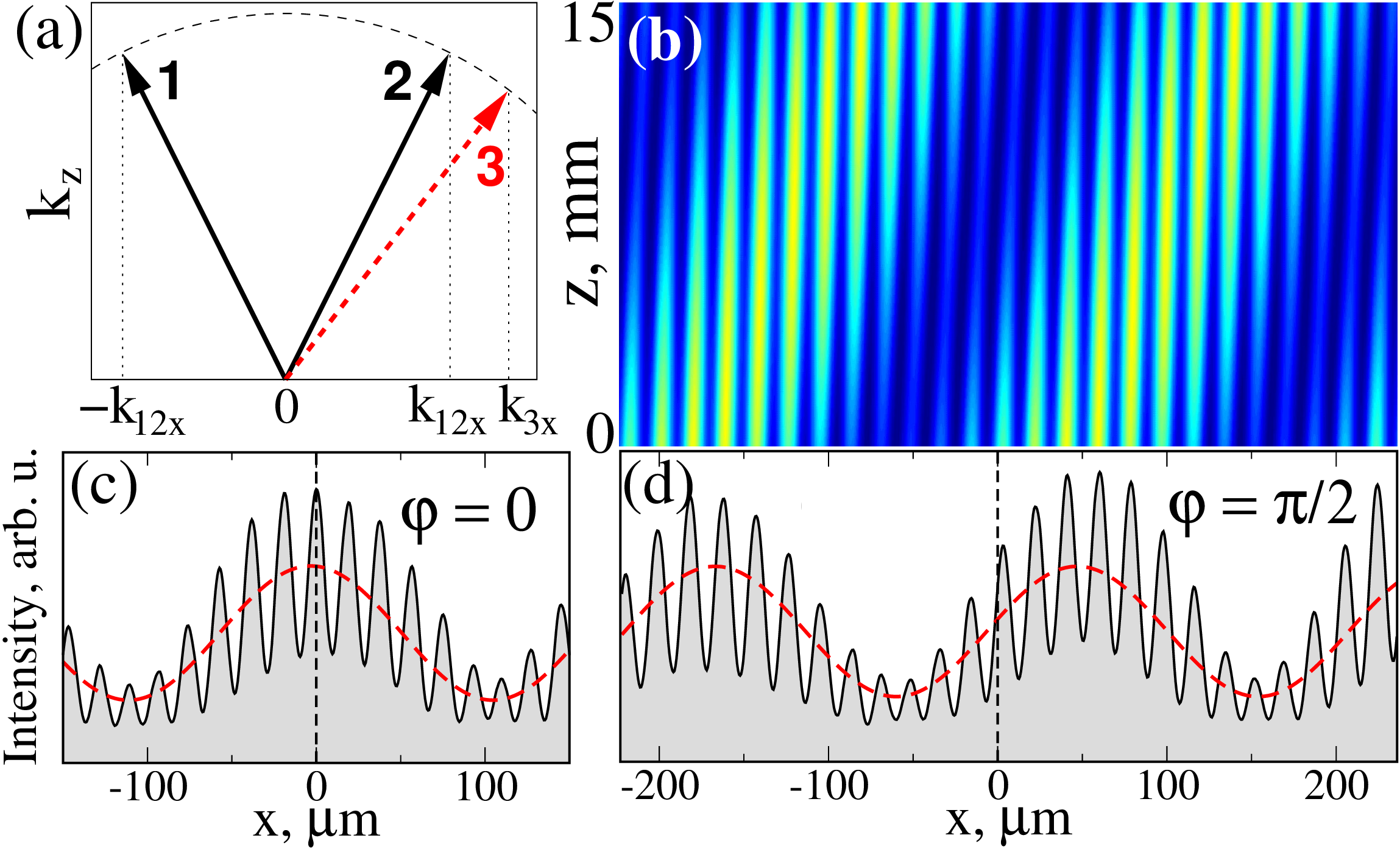}{modul_lattice}{ 
(color online)~(a)~Schematic of the $k$-vector configuration in a modulated optical lattice. Beams {\sf 1} and {\sf 2} define a straight lattice, while the symmetry breaking control beam {\sf 3} introduces transverse and longitudinal lattice modulation, as shown in (b) for the case $k_{3x} = 1.18 k_{12x}$ and $I_3 = 4 I_{12}$. Bottom: corresponding experimental profiles of modulated lattice at the crystal input ($z=0$), when maximum of the broad transverse modulation (red-dashed) is (c)~aligned with probe beam input ($x=0$), and (d)~offset by a quarter period. }

Similar to Refs.~\cite{Kartashov:2005-1378:OL, Garanovich:2005-5704:OE}, we consider a modulated lattice created by two waves with equal amplitudes $A_{12}$ and opposite inclination angles defined by the transverse wavenumbers $k_{12x}$ and $-k_{12x}$, and an additional third wave with amplitude $A_3$ and wavenumber $k_{3x}$ [see Fig.~\rpict{modul_lattice}(a)]. In numerical simulations, we normalize the transverse $x$ and propagation $z$ coordinates to the characteristic scales $x_s$ and $z_s$, respectively. Then, the three-wave interference pattern is $I_p(x,z)=|A_L|^2$, where $A_L = A_3 \exp[ i \beta_3 z + i k_{3x} x - i \varphi] + 2 A_{12} \exp( i \beta_{12} z ) \cos( k_{12x} x)$, $\varphi$ is the relative phase between the third wave and the other two waves, and the propagation constants $\beta_j = D k_j^2$ define the longitudinal wavevector components. Here $D = z_s \lambda / (4 \pi n_0 x_s^2)$ is the diffraction coefficient, $n_0$ is the average refractive index of the medium, and $\lambda$ is the wavelength in vacuum.
Figure~\rpict{modul_lattice}(b) shows an example of the modulated lattice. Changing the phase of the third beam $\varphi$ causes the whole lattice pattern to shift in both transverse [cf. Fig.~\rpict{modul_lattice}(c) and Fig.~\rpict{modul_lattice}(d)] and longitudinal directions. The lattice also depends on the inclination angle and power of the third beam, characterized by the parameters $k_{3x}/k_{12x}$, and  $I_3/I_{12}$, where $I_3=|A_3|^2$ and $I_{12}=|A_{12}|^2$. 
We model the propagation of an extraordinarily-polarized beam 
by a parabolic equation for the normalized beam envelope $E(x,z)$,
$i {\partial E}/{\partial z} + D {\partial^2 E}/{\partial x^2}+ {\cal F}(x,z,|E|^2) E= 0$,
where the function ${\cal F}$ is proportional to the optically-induced change of refractive index. Due to the strong electro-optic anisotropy of certain photorefractive crystals such as SBN, this term almost vanishes for the lattice-writing beams polarized orthogonal to the {\em c}-axis of the crystal because of a very small effective nonlinear coefficient~\cite{Efremidis:2002-46602:PRE}, while the extraordinarily-polarized probe beam experiences a highly nonlinear evolution~\cite{Efremidis:2002-46602:PRE, Fleischer:2003-23902:PRL, Neshev:2003-710:OL, Sukhorukov:2004-93901:PRL} with ${\cal F}( x, z, |E|^2) = - \gamma (I_b + I_p(x, z) + |E|^2)^{-1}$, where $I_b$ is the constant dark irradiance, $I_p(x, z)$ is the lattice interference pattern, and $\gamma$ is the nonlinear coefficient proportional to the applied DC field. To match our experiments, we use the following parameters: $\lambda=0.532\mu$m, $n_0=2.35$, $x_s=1\mu$m, $z_s=1$mm, $I_b=1$, $\gamma=2.36$, and $d=20.0\mu$m is the period of the non-modulated lattice (for $A_3=0$). The crystal length is $L=15$mm.

In experiment, we create a modulated optical lattice by interfering three ordinarily-polarized broad laser beams from a frequency-doubled Nd:YVO$_4$ cw laser at a wavelength 532~nm inside a $15\times5\times5$~mm SBN:60 crystal. Applying an external electric field of $2.2$~kV, we produce a refractive index modulation and control the saturation by homogeneously illuminating the crystal with white light. An extraordinarily-polarized Gaussian probe beam with a full width at half maximum (FWHM) of $25\mu$m  (along the $x$ direction and extended in $y$) is launched into the crystal, parallel to the $z$ axis ($k_x=0$). The front and back faces of the crystal are imaged onto a CCD camera to capture the probe beam intensity distributions. A second imaging system is used to monitor the input position of lattice fringes and probe beam, the latter of which is fixed at $x=0$.

\pict{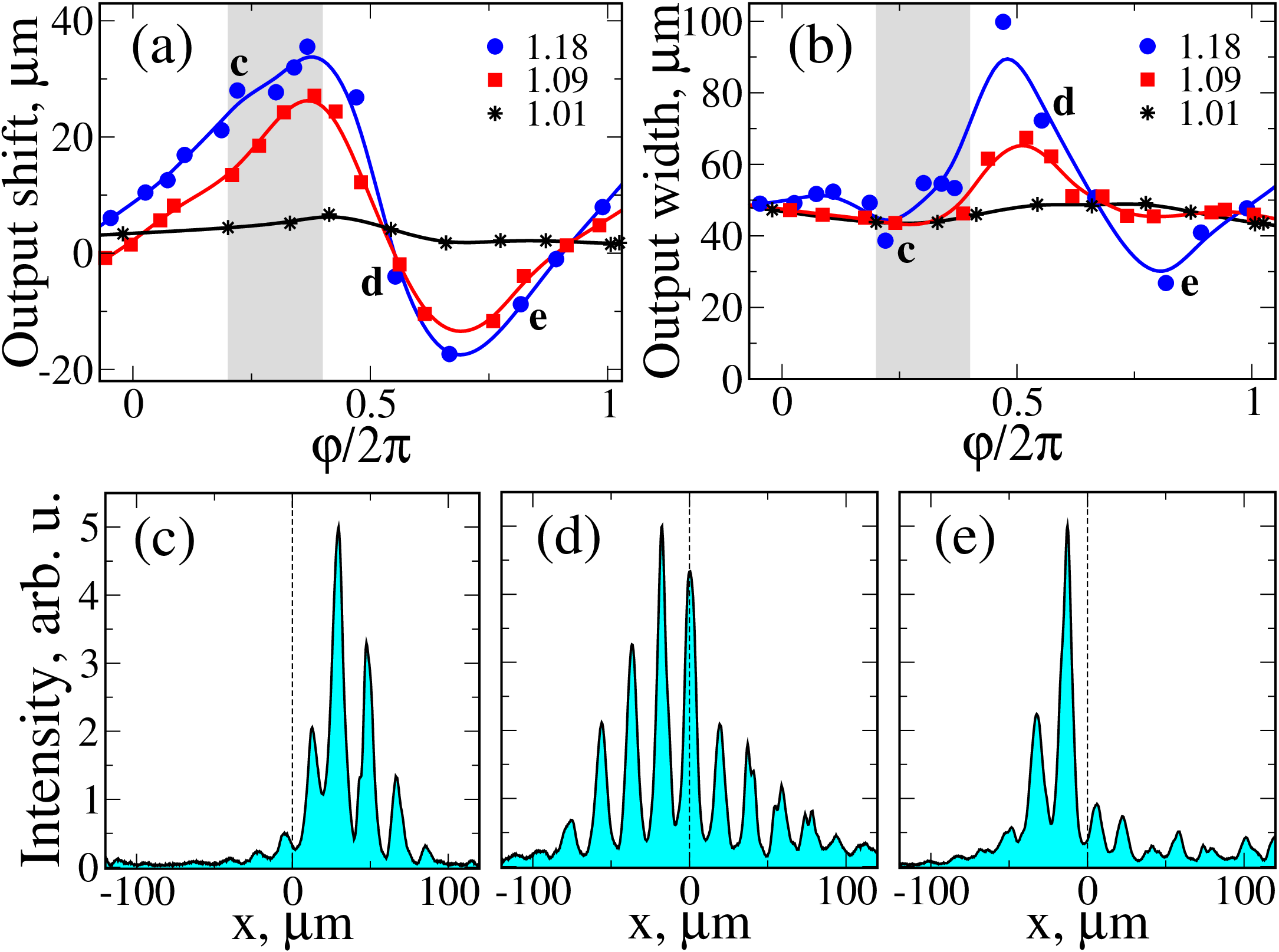}{shift_phase}{ (color online)~(a,b) Measured shift
and width of a linear probe beam output vs. the phase of the
modulating lattice beam for three different values of
$k_{3x}/k_{12x}$ for $I_3=4I_{12}$. Shading marks the region in
which the figure of merit is maximized. (c-e) Examples of
measured output profiles corresponding to the points (c,d,e) in
the plots (a,b). }

A transverse lattice symmetry is recovered when $k_{3x}=0$ or $k_{3x}=\pm k_{12x}$. In all other cases, the modulation is asymmetric, and steering of the normally incident probe beam becomes possible. We determine experimentally that, due to the finite trapping strength of the lattice, steering without strong beam reshaping or break-up is possible for a limited parameter range $0.8<k_{3x}/k_{12x}<1.2$. Below we discuss in further detail the case $k_{3x}>k_{12x}$, noting that the steering behavior in the complementary regime $k_{3x}<k_{12x}$ is similar.

First, we characterize the effect of the modulated lattice geometry on the propagation of a low-power ($\sim25nW$) probe beam in the linear regime for three different angles of the modulating beam, in the case of strong lattice modulation, $I_3=4I_{12}$. In Fig.~\rpict{shift_phase}(a) we plot the shift of the beam center of mass vs. the modulating beam phase $\varphi$ [see Fig.~\rpict{modul_lattice}(c,d)]. Solid lines represent
smoothing spline fitting to the data points [experimental
uncertainty in the vertical direction is approximately $10\mu$m in
Fig.~\rpict{shift_phase}(a), and up to $20\mu$m in Fig.~\rpict{shift_phase}(b)].
The phase $\varphi$ is adjusted by passing the third beam through a
thin glass plate with a variable tilt. For $k_{3x}=1.01k_{12x}$ the
beam shift is virtually zero throughout the entire phase scan [stars
in Fig.~\rpict{shift_phase}(a)]. The insensitivity to $\varphi$ results
from the lattice being fully symmetric when beams {\sf 2}
and {\sf 3} in Fig.~\rpict{modul_lattice}(a) are parallel ($k_{3x}=k_{12x}$).

\pict{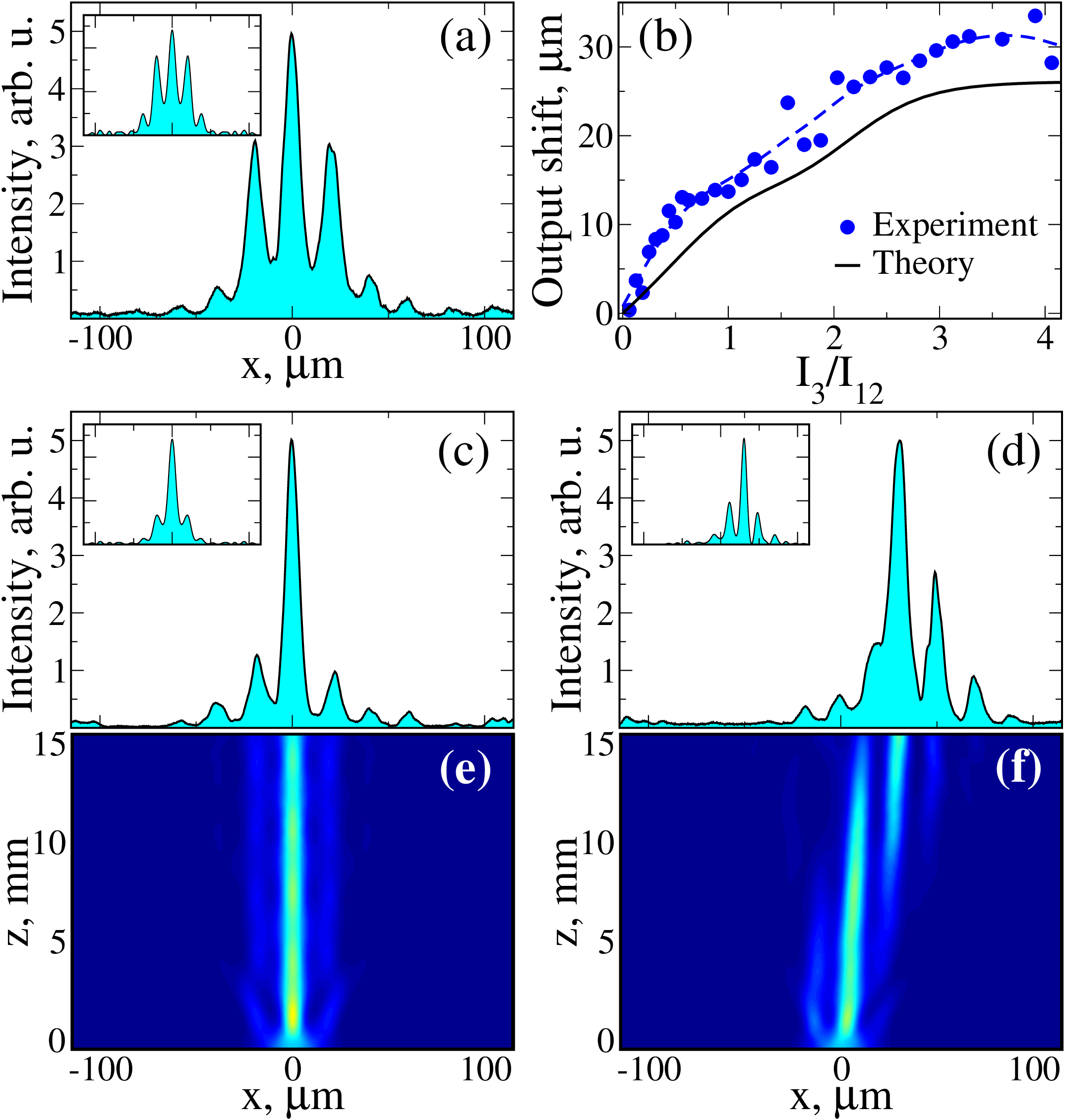}{shift_nonlinear}{ (color online)~ (a)~Experimental and
theoretical (inset) linear output in a straight lattice ($I_3=0$).
(b)~Shift of the nonlinear probe beam output vs. the modulating beam
power for $k_{3x} = 1.18 k_{12x}$ and $\varphi/2\pi=0.22$. (c,d)~Experimental and
theoretical (inset) nonlinear output in straight and
modulated lattices for $I_3=0$ and $I_3=4I_{12}$, respectively, and
$k_{3x} = 1.18 k_{12x}$. (e,f)~Numerical simulations of the
longitudinal propagation. }

On the other hand, as the angle of the third lattice-forming wave is increased, the lattice becomes asymmetrically modulated, and the beam shifts to one or the other side [squares and circles in Fig.~\rpict{shift_phase}(a)], depending strongly on the value of $\varphi$ (similar behavior was observed for $k_{3x}<k_{12}$). This proves that not only the local asymmetric distortion of the lattice, which in this case tends to
shift the beam towards positive $x$ [see Fig.~\rpict{modul_lattice}(b)], but also the broader effective modulation geometry plays an important
role for the beam propagation dynamics.
We further note that in Fig.~\rpict{shift_phase}(a) the beam shift is not symmetric with respect to $x=0$, and it is strongest in the region $0<\varphi<\pi$ where the effects of local and broad lattice modulation pull the beam in the same direction.

Figure~\rpict{shift_phase}(b) maps the corresponding output beam width
(FWHM of Gaussian fit) as a function of $\varphi$. Again, the case
$k_{3x}=1.01k_{12x}$ proves to be relatively insensitive to the
phase shift, whereas for larger angles, substantial beam broadening
is observed close to $\varphi=\pi$. For $k_{3x}=1.18k_{12x}$, the
maximum output beam width greatly exceeds that of a diffracting beam
in the absence of a lattice ($57\mu$m), and the observed broadening
must be attributed to the geometry of the modulated
lattice, and not solely to the decreased contrast of the lattice at
$\varphi=\pi$ [see Fig.~\rpict{modul_lattice}]. 
We note that when the phase $\varphi$ is scanned from zero to $2\pi$, the local input beam excitation symmetry changes from on-site to off-site and back, and this may in principle lead to additional beam steering~\cite{Morandotti:1999-2726:PRL}. However, we verified that under our experimental conditions the contributions to center of mass shift as well as beam broadening due to this effect are negligible ($<3\mu$m in both cases). Examples of output beam profiles corresponding to points c,~d~and~e in Figs.~\rpict{shift_phase}(a,b) are shown in Figs.~\rpict{shift_phase}(c-e).

The angle and phase of the modulating beam are important control
parameters, and for practical applications it is
necessary to realize an optimal balance between 
the effects of beam shift and broadening. To
characterize the steering performance, we quantify the spatial resolution by 
a figure of merit $F=\left|\Delta
x\right|/W$, where $\Delta x$ and $W$ are the shift and the width of the
output beam, respectively. In Figs.~\rpict{shift_phase}(a,b) gray shading
marks the region in which the figure of merit exceeds $0.5$ and
climbs to a maximum of $0.6$ for the largest modulation angle
$k_{3x} = 1.18 k_{12x}$. Focusing now on this case [see
Fig.~\rpict{shift_phase}(c)], we show in Fig.~\rpict{shift_nonlinear} that increasing the power of the probe beam to $1.5\mu$W leads to strong
self-focusing and enhanced beam localization [Fig.~\rpict{shift_nonlinear}(c)] while preserving a
large beam shift [Fig.~\rpict{shift_nonlinear}(d)]. As a result, the figure
of merit is increased by approximately a factor of two compared to
the linear case, thus exceeding unity. Figure~\rpict{shift_nonlinear}(a) shows,
for comparison, the broader low power output profile in a regular
straight lattice ($I_3=0$). In all cases the experimental
observations match our numerical simulations, shown in
Fig.~\rpict{shift_nonlinear} as the beam profile insets, in panels (a,c,d),
and the top view of the propagation dynamics, in panels (e,f).
In Fig.~\rpict{shift_nonlinear}(b) we trace the experimental (circles) and
theoretical (solid line) nonlinear beam shift as a function of the
lattice modulation power. 
We find that the beam shift gradually increases and, in experiment, saturates at approximately $\Delta x=30\mu$m for $I_3/I_{12}>3$. The difference between theory and experiment in Fig.~\rpict{shift_nonlinear}(b) may be attributed to a small self-induced drift of the strongly localized beam, that was not taken into account in simulations.

In conclusion, we have demonstrated experimentally all-optical
steering of nonlinear self-localized beams in modulated
optically-induced photonic lattices. We have revealed the key
features associated with the beam steering effect and described how
it depends on inclination angle, phase, and power of the
modulating lattice beam. These results allowed us to optimize the
steering performance and achieve high spatial resolution with a
figure of merit exceeding unity.

\end{sloppy}
\end{document}